\documentclass[12pt,preprint]{aastex}
\usepackage{graphicx,picins,wrapfig}           
\usepackage[english]{babel}
\usepackage[T1]{fontenc}
\usepackage[ansinew]{inputenc}
\usepackage{placeins}
\usepackage{ae}
\usepackage{amsmath,amssymb,amstext,amscd,txfonts,bm}
\usepackage{natbib}
\usepackage{times}
\usepackage{lettrine}
\usepackage{fancyhdr}           

\newcommand\tc[2]{$\begin{cases}\;#1\\\;#2\end{cases}$}
\newcommand\mt[2]{$\begin{matrix}#1\\#2\end{matrix}$}
\newcommand{\myemail}{klement@mpia.de}
\newcommand{\fuchsemail}{fuchs@ari.uni-heidelberg.de}
\defcitealias{kle08}{KFR08}
\defcitealias{bre10}{B+10}
\begin{document}

\title{Classification of field dwarfs and giants in RAVE and its use in stellar stream detection}

\author{R.~J. Klement\altaffilmark{1}, C.~A.~L. Bailer-Jones\altaffilmark{1}, B. Fuchs\altaffilmark{2}, H.-W. Rix\altaffilmark{1} and K.~W. Smith\altaffilmark{1}}

\altaffiltext{1}{Max-Planck-Institut f\"ur Astronomie, K\"onigstuhl 17, D-69117 Heidelberg; \myemail}
\altaffiltext{2}{Astronomisches Rechen-Institut am Zentrum f\"ur Astronomie Heidelberg, M\"onchhofstraße 12-14, D-69120 Heidelberg; \fuchsemail}

\begin{abstract}
Samples of bright stars, as they emerge from surveys such as RAVE, contain comparable fractions of dwarf and giant stars. An efficient separation of these two luminosity classes is therefore important, especially for studies in which distances are estimated through photometric parallax relations. We use the available spectroscopic $\log g$ estimates from the second RAVE data release (DR2) to assign each star a probability for being a dwarf or subgiant/\,giant based on mixture model fits to the $\log g$ distribution in different color bins. We further attempt to use these stars as a labeled training set in order to classify stars which lack $\log g$ estimates into dwarfs and giants with a SVM algorithm. We assess the performance of this classification against different choices of the input feature vector. In particular, we use different combinations of reduced proper motions, 2MASS $JHK$, DENIS $IJK$ and USNO-B $B2R2$ apparent magnitudes. Our study shows that -- for our color ranges -- the infrared bands alone provide no relevant information to separate dwarfs and giants. Even when optical bands and reduced proper motions are added, the fraction of true giants classified as dwarfs (the contamination) remains above 20\%. 

Using only the dwarfs with available spectroscopic $\log g$ and distance estimates (the latter from Breddels et al. 2010), we then repeat the stream search by Klement, Fuchs \& Rix (2008, KFR08), which assumed all stars were dwarfs and claimed the discovery of a new stellar stream at $V\approx-160$ km s$^{-1}$ in a sample of 7015 stars from RAVE DR1. The existence of the KFR08 stream has been supported by two recent studies using other independent datasets. Our re-analysis of the pure DR2 dwarf sample exhibits an overdensity of 5 stars at the phase-space position of the KFR08 stream, with a metallicity distribution that appears inconsistent with that of stars at comparably low rotational velocities. Compared to several smooth Milky Way models, the mean standardized deviation of the KFR08 stream is only marginal at $1.6\pm0.4$. Our data therefore do not allow to draw definite conclusions about its existence, but future RAVE data releases and other (best spectroscopic) surveys are going to help in resolving this issue.
\end{abstract}


\section{Introduction}
In the last decade we have witnessed a tremendous increase of astronomical data due to large automatic surveys such as 2MASS \citep{skr06}, SDSS/SEGUE \citep{york00,yan09} or RAVE \citep{stei06}. The data produced by such surveys -- apparent magnitudes and/or spectra -- need further processing to determine the type and properties of the observed sources. This step often requires automatic algorithms that search for regularities in the data and use this information to classify the data into different categories (\textit{classification}) or to determine the values of certain parameters (\textit{regression}). \textit{Machine learning} techniques are ideally suited for such purposes, and their applications have dramatically grown in astronomy. For example, \citet{rich04} used the location of 16,713 spectroscopically classified quasars in SDSS color-space to identify 100,565 quasar candidates in a sample of SDSS DR1 point sources based on kernel density estimation. Similarly, using neural networks trained on spectroscopically pre-classified high-redshift SDSS quasars with optical photometry and radio data, \citet{car08} obtained 58 high-redshift quasar candidates in data taken from the FIRST survey; spectroscopic follow-up observations confirmed that 60\% of the candidates were true high-redshift quasars. A study from the same year by \citet{gao08} compared the performance of Support Vector Machines and $k$-dimensional trees in the classification of quasars using different combinations of SDSS and 2MASS colors and magnitudes. They concluded that both algorithms yield high efficiency; SVMs have slightly higher accuracy but need also more computation time. An example of automated regression can be found in \citet{refi07}; these authors used non-linear regression models trained either on pre-classified observed SDSS/SEGUE spectra or synthetic spectra to estimate the stellar atmospheric parameters $T_\text{eff}$, $\log g$ and [Fe/H]. 

The main goal of the present study is to identify dwarfs in the second RAVE public data release \citep[DR2,][]{zwi08} in order to build a sample with minimal contamination from subgiants and giants. Only for such a sample, reliable distance estimates are possible, either from photometric parallaxes (not used in this paper) or the recently published RAVE distances which were derived from stellar models \citep[][hereafter B+10]{bre10}. DR2 includes the first data release \citep[DR1,][]{stei06}, which was used by \citet[hereafter KFR08]{kle08} to claim the discovery of a new stellar stream in the solar neighborhood. Their analysis was based on the assumption that the vast majority of RAVE stars are main sequence stars; as we describe in the next section, this assumption is questionable and deserves deeper investigation. We therefore try to redo the analysis from \citetalias{kle08} on a preferably large and pure dwarf sample and hope to recover a signal of the new stream.

The reason that DR1, and also part of DR2, contains no estimate for the stellar parameter $\log g$, which would allow a discrimination between dwarf and giant, is that the corresponding second-order spectra are contaminated by blue light, because a later installed blue-blocking filter (Schott OG513) was not used during the first year. However, part of DR2 contains estimates for $\log g$, and we use these to build a training set of stars for automatic classification of the others based on features like photometry and proper motions.

This paper is organized as follows: In \S~\ref{sec:s2}, we show that early and late-type stars with available $\log g$ estimates naturally fall into two or three distinct classes, respectively, which we naively interpret as dwarfs, subgiants and giants. We can thus assign class labels to these stars and use them to investigate whether it is possible to classify stars into dwarfs and ``others'' based on photometry and/or proper motions alone. This is the subject of \S~\ref{sec:s3}, where we show that such a classification, performed by a Support Vector Machine (SVM) algorithm, can not yield an output dwarf sample in which the contamination from giants is below $\sim20\%$. In \S~\ref{sec:s4}, we therefore use only the dwarfs that have been labelled on the basis of their spectroscopic $\log g$ estimates together with distances from \citetalias{bre10} to search for stellar streams with our method from \citetalias{kle08}. We show that we are able to recover a signal of the proposed new stream, which is however not highly significant. Our results are summerized in \S~\ref{sec:s5}

\section{The Mixture Model Classification\label{sec:s2}}
\subsection{The Data\label{sec:s21}}
Based on photometric parallaxes calibrated with \textit{Hipparcos} data, in \citetalias{kle08} we searched for overdensities in the phase space of nearby ($d<500$ pc) stars drawn from DR1 and detected a feature located at an azimuthal velocity of $V\approx-160$ km s$^{-1}$ with respect to the Local Standard of Rest. This feature contained 5 stars in a region of phase space that proved to be significantly clumped compared to a smooth Milky Way model ($\sigma>2$), while another 14 stars just outside this region probably also belonged to this stellar ``stream''. This was based on the assumption that ``the vast majority of RAVE stars are main sequence stars, because giants in the RAVE magnitude range would lie so far away, that the star density is substantially decreased.''

However, \citet{sea08} have estimated that only $\sim5\%$ of the RAVE stars are K-M dwarfs, while K-M giants account for $\sim44\%$. We therefore wish to critically check our assumption by using the stars from DR2 with spectroscopic $\log g$ estimates to build a new sample of dwarfs with minimal contamination from the other luminosity classes. The $\log g$ values were derived by a penalized $\chi^2$ method to construct a synthetic spectrum that matches the observed spectrum \citep{zwi08}. The observed spectrum is modelled as a weighted sum of Kurucz template spectra with known parameters and it is assumed that the parameters follow the same weight relation. From a comparison to external datasets, the typical $\log g$ uncertainties for stars with a typical S/N ratio of 40 range from $\sim0.15$ dex for dwarfs with $T_\text{eff}=9000$ K to $\sim0.7$ dex for cooler dwarfs with $T_\text{eff}=6000$ K \citep[][Fig.~19]{zwi08}. 

Recently, \citetalias{bre10} published distance estimates for 16,075 stars in the DR2 catalog with typical relative uncertainties in the range 30\%--50\%. These distances were derived by fitting the observed $\log g$, $T_\text{eff}$, [M/H] and $(J-K)$ values of each star to one out of 600 theoretical isochrones from the Y$^2$ \citep[Yonsei-Yale,][]{dem04} models, in this way obtaining absolute magnitudes in the $J$ band. Uncertainties on the distances have been computed out of the probability distribution function which is obtained by drawing a sample of 5000 Monte-Carlo stars from a 4D Gaussian centered on the model isochrone with standard deviation in each parameter equal to the uncertainty of the corresponding observed parameter of the true star, and again finding the best-matching isochrone for each of the 5000 Monte Carlo stars. By comparing their distances to those of main sequence stars from the \textit{Hipparcos} catalog and giants in the open cluster M67, \citetalias{bre10} concluded that their distances for the main sequence stars are reliable, with no systematic errors exceeding $\sim10\%$, whereas the giants were systematically offset from the expected M67 isochrone. This again highlights our wish to build a pure sample of dwarfs based on the available $\log g$ estimates.

\subsection{Creating a labeled sample of Dwarfs and Giants based on mixture models\label{sec:s22}}
Before we start investigating the $\log g$ distribution, we take the following considerations into account: first, the gravity parameter of a given luminosity class changes with spectral type (color). For example, $\log g$ is expected to be lower for G dwarfs than for M dwarfs (of the same metallicity). Second, reddening of stars leads to a less clear separation of dwarfs from other luminosity classes in the $\log g$ distribution, because a reddened giant will have a higher $\log g$ value than an unreddened one with the same color, while for dwarfs the opposite is true (although in the RAVE magnitude range reddening will primarily effect the more distant giants). Third, the dwarf-giant separation will get better with increasing color, because the number of turn-off stars and subgiants decreases. Fourth, the expected small fraction of dwarfs \citep{sea08} in combination with their higher $\log g$ uncertainties might hinder a clear separation. Finally, the $\log g$ distribution has been shown to vary with galactic latitude $b$ \citep[Fig.~22]{zwi08}.

Keeping these considerations in mind, we prepare the DR2 sample for our purposes in the following way: of the 51,829 objects contained in DR2, only 49,327 objects are unique, the rest being repeated observations of single stars \citep[Table~1]{zwi08}. If a star has been observed multiple times, we assign it the mean radial velocity of these observations. Next, we remove all stars without photometry in all of the 2MASS $JHK$ bands, because these bands are the most comprehensive ones and least affected by extinction. We further decrease any redding effects by keeping only stars with latitudes $b>20^\circ$. This step has the additional benefit that the number of giants goes down due to their decreasing space density in the halo \citep[see also][Fig.~23]{zwi08}. Our cuts leave 48,335 individual stars of which a subset of 20,080 have estimates for $\log g$. To account for the varying fraction of dwarfs and subgiants/giants with galactic latitude, we derive $\log g$ distributions separately for stars with $20^\circ<\vert b\vert\leq40^\circ$ and $40^\circ<\vert b\vert\leq90^\circ$, respectively.

\begin{figure}[ht]
\epsscale{1.0}
\plotone{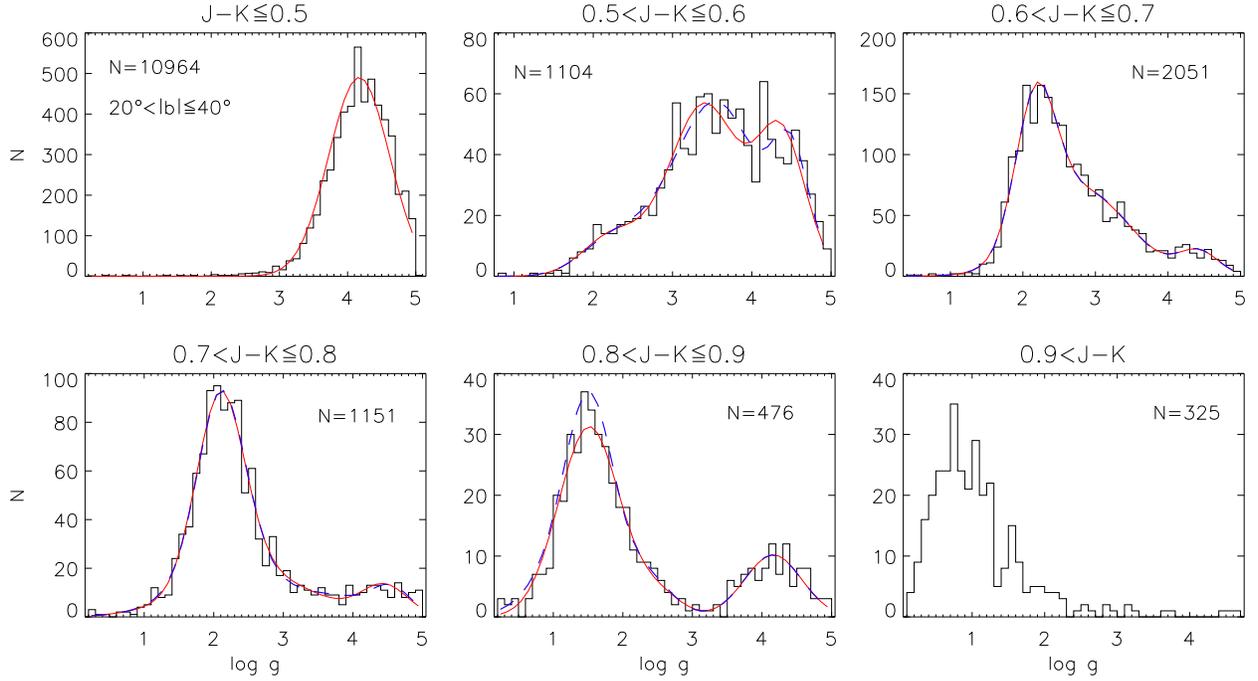}
\caption{$\log g$ histograms for stars at galactic latitudes $20^\circ<\vert b\vert\leq40^\circ$ and with $(J-K)\leq0.5$ (\emph{upper left panel}) and five bins of stars with $(J-K)>0.5$ (\emph{other panels}). The curves show mixture model fits to what we consider a dwarf and a ``giant'' population. For the early type stars, the distribution can be fitted by a single Gaussian, while for the late type stars three components are fit (red curve). The blue dashed curve gives a fit with four components and looks almost indistinguishable to the three-component fit. The $\log g$ distributions do not extend beyond 5.0 dex, which is the grid limit of the spectral libraries for RAVE. \label{fig:f1}}
\end{figure}

\begin{figure}[ht]
\epsscale{1.0}
\plotone{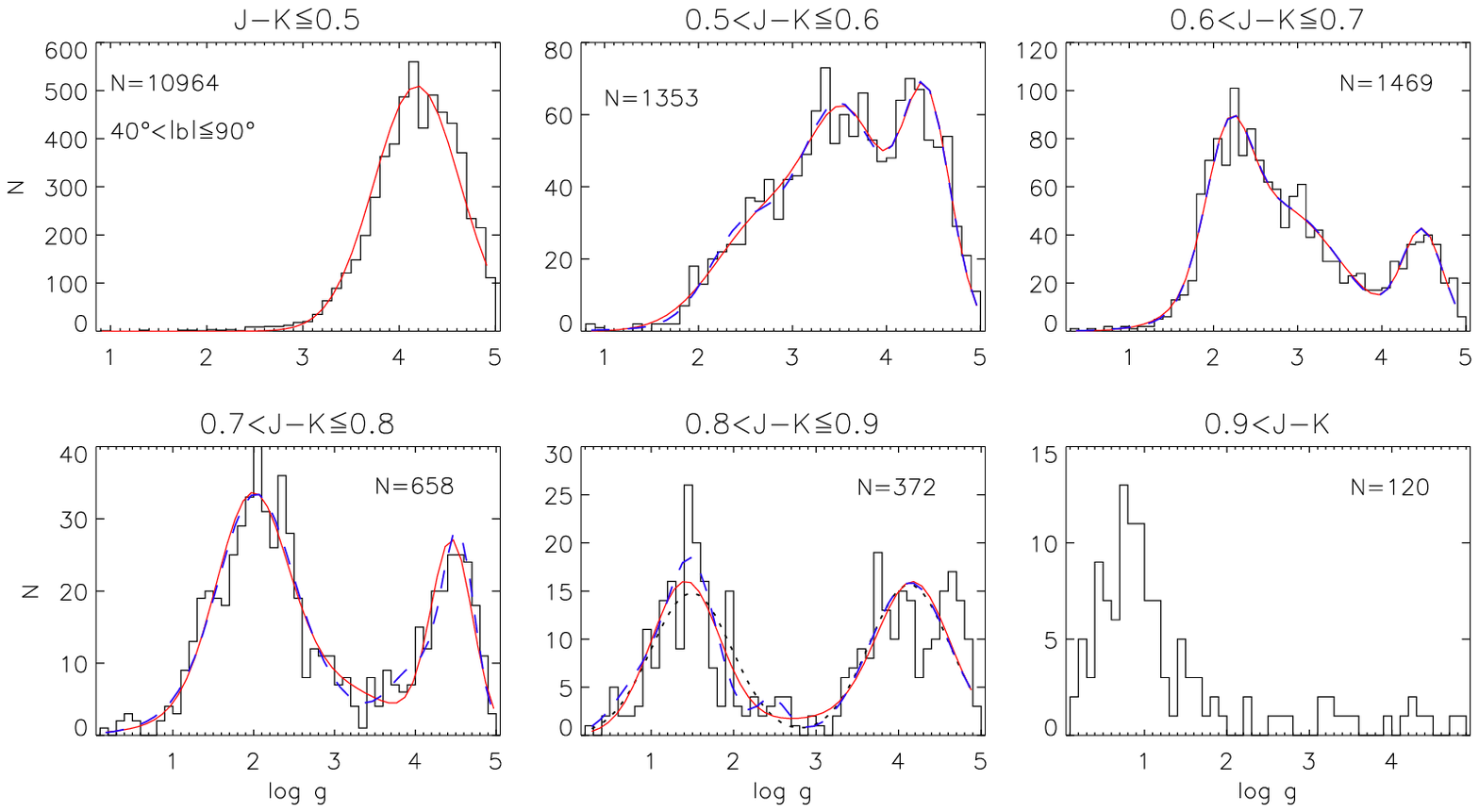}
\caption{Same as Figure~\ref{fig:f1}, but now for stars with $40^\circ<\vert b\vert\leq90^\circ$. For the color bin $0.8<J-K\leq 0.9$ we additionally show a two-component mixture model fit (black dotted line) since subgiants should be negligible in this color range.  \label{fig:f2}}
\end{figure}

These are displayed in Figures~\ref{fig:f1} and \ref{fig:f2}, where we have separated stars with $(J-K)\leq0.5$ (the ``early-type'' subsample, upper left panel) from those with $(J-K)>0.5$ (the ``late type'' subsample, which is further binned into the other panels). According to synthetic 2MASS colors calculated by \citet[][Table~3]{cov07}, a solar-metallicity K0 star has $(J-K)=0.54$. Hence the color cut at $(J-K)=0.5$ roughly separates K and M stars from earlier types. From Figures~\ref{fig:f1} and \ref{fig:f2} it is also apparent that it divides our DR2 sample into two parts with respect to the fraction of giants: in the early type subsample, the contamination from giants seems to be negligible; the $\log g$-distribution can be fitted by a single Gaussian with mean $\mu=4.2$ dex and standard deviation $\sigma=0.44$ dex, independent of latitude $b$. In this subsample, the fraction of stars that lie outside of the $2\sigma$-region of the Gaussian (at $\log g\leq3.09$, which we decide to attribute to giants) is less than 3.9\%. The Gaussian most likely contains dwarfs as well as turnoff-stars and subgiants, because the transitions in this color range are continuous. In the late-type subsample, on the other hand, we clearly have more than one population of stars, and the $\log g$ distributions change with color.\footnote{The $\log g$ distribution in the early type subsample changes much less prominent with color, so that we marginalized over colors.} We use an EM-algorithm \citep[see e.g.][Section~8.5]{has08} to fit the $\log g$-distributions with mixture models of multiple Gaussians, in this way obtaining probability densities without using class labels (unsupervised classification). We found that in general, three Gaussians fit the observed distributions well, while adding more components can result in overfitting of the data.\footnote{The fit to the high-latitude stars in the color range $0.8<J-K\leq0.9$ is not so easy to interpret, because we found that the solution of the EM algorithm in this color range is not unique: if we choose initial mean values of $\log g\gtrsim3$ for two of the Gaussians, the algorithm converges to a different solution, which fits the distribution at $\log g\gtrsim3$ with two narrow Gaussians caused by the ``gap'' at $\log g\sim4.2$. However, stars in this color range should not have yet developed into subgiants, so that two narrow Gaussians are not only statistically, but also physically unrealistic; a better explanation for the gap would be the large uncertainties in the $\log g$ values for red stars. The lack of subgiants should allow a two-component fit and indeed we find that a two-component fit is almost indistinguishable from the three-component one (black dotted line in Fig.~\ref{fig:f2}). We nevertheless adopt the three-component fit to be consistent with the other color bins, knowing that the separation of dwarfs will be very similar.} This shows that we can generally decompose stars in the different color bins into three classes, which we can think of being dwarfs, subgiants/intermediates and giants. An exeption is the color range $(J-K)>0.9$, where we do not fit the data, because in this color range the signature of the dwarf population is washed out due to their small number and large $\log g$ uncertainties. We therefore do not use stars with $(J-K)>0.9$ \citep[M4/M5 and later,][]{cov07} in our further analysis, also because these stars fall out of the range of the synthetic spectra which were used to obtain stellar parameters \citep[][Sect.~3.2.]{zwi08}. 

The parameters of the different mixture models are given in Table~\ref{tab:t1}. As expected, the weighting of the median Gaussian component associated with the contribution from subgiants decreases with increasing color, from 0.57 at $0.5<J-K\leq0.6$
to 0.06 at $0.8<J-K\leq0.9$ for the low latitude sources. Also, the fraction of the third component associated with dwarfs is higher at high galactic latitudes for all color bins except $0.5<J-K\leq0.6$, where it is roughly constant at $\sim30\%$. This trend has already been described by \citet{zwi08} and is attributed to the fact that the RAVE giants trace a systematically more distant population and already belong to the scarcely populated halo at high latitudes. There is a clear trend of the means of all three components to shift to lower values with increasing color, e.g. from $\log g=4.4$ at $0.5<J-K\leq0.6$ to $\log g=4.2$ at $0.8<J-K\leq0.9$ for the dwarfs, reflecing the decreasing stellar masses of giants, subgiants and dwarfs for later types. This is accompanied by a trend of the standard deviations to increase, which is caused by the fact that the $\log g$ uncertainties increase with color. The color bin $0.8<J-K\leq0.9$ allows the clearest separation between dwarfs and giants due to the very few subgiants, expressed through a relatively large standard deviation of the dwarfish Gaussian component.

\begin{deluxetable}{c|c|c|c|c}
\tablewidth{0pt}
\tablecaption{Parameters of the mixture models\label{tab:t1}}
\tabletypesize{\footnotesize}
\tablehead{
\colhead{$(J-K)$}  & \colhead{$\bm\pi$} &  \colhead{$\bm\mu$} & \colhead{$\bm\sigma$} & \colhead{$\log g_t$} \\
\colhead{mag}  & \colhead{ } &  \colhead{(dex)} & \colhead{(dex)} & \colhead{(dex)}
} 
\startdata
			0.5...0.6  & \mt{(0.12,0.57,0.31)}{(0.31,0.41,0.28)} & \mt{(2.25,3.40,4.38)}{(2.72,3.61,4.43)} & \mt{(0.38,0.45,0.30)}{(0.54,0.41,0.25)} & \mt{4.02}{4.11}\\
			0.6...0.7  & \mt{(0.39,0.55,0.06)}{(0.25,0.59,0.16)} & \mt{(2.17,2.82,4.43)}{(2.18,2.79,4.48)} & \mt{(0.27,0.64,0.27)}{(0.26,0.68,0.24)} & \mt{4.10}{4.07}\\
			0.7...0.8  & \mt{(0.57,0.34,0.09)}{(0.44,0.31,0.25)} & \mt{(2.10,2.51,4.46)}{(1.97,2.54,4.45)} & \mt{(0.34,0.86,0.33)}{(0.43,0.97,0.26)} & \mt{3.98}{3.92}\\
			0.8...0.9  & \mt{(0.73,0.06,0.21)}{(0.45,0.08,0.47)} & \mt{(1.54,2.66,4.20)}{(1.41,2.89,4.18)} & \mt{(0.46,0.73,0.38)}{(0.42,0.76,0.45)} & \mt{3.50}{3.21}\\
\enddata
\tablecomments{In each row, the upper part corresponds to the stars with $20^\circ<\vert b\vert\leq40^\circ$, the lower part to those with $40^\circ<\vert b\vert\leq90^\circ$. The columns give the valid color range for the mixture model, the vector of mixing coefficients $\bm\pi$, the vectors of means $\bm\mu$ and standard deviations $\bm\sigma$ of the Gaussian components, and $\log g_t$, the threshold value where the probability of a star for being a dwarf gets greater than 0.5. The latter is computed by assuming that the dwarfs are described by the third Gaussian at the right of the distribution.} 
\end{deluxetable}

From Figures~\ref{fig:f1} and \ref{fig:f2} and Table~\ref{tab:t1} we see that the three-component mixture models fit the observed $\log g$-distributions reasonably well and -- as a result -- predict the late-type dwarfs to be located around $\log g\approx4.4$. Adopting this model, we calculate a star's probability for being a dwarf as
\begin{equation}\label{eq:e1}
	\wp\bigl(\text{dwarf}\,\vert\,\log g, (J-K),\vert b\vert\bigr)=\frac{\pi_3(J-K)\mathcal{N}\bigl(x_i\,\vert\,\mu_3(J-K,\vert b\vert),\sigma_3(J-K,\vert b\vert)\bigr)}{\sum_{k=1}^3\pi_k(J-K,\vert b\vert)\mathcal{N}\bigl(x_i\,\vert\,\mu_k(J-K,\vert b\vert),\sigma_k(J-K,\vert b\vert)\bigr)},
\end{equation}
where the index 3 refers to the Gaussian component on the right side of the $\log g$-distribution, that we choose to interpret as representing the dwarfs, and the parameters depend on the color $(J-K)$ and the absolute value of latitude $\vert b\vert$ given in Table~\ref{tab:t1}. 

In what follows, we simply combine subgiants, giants and supergiants into a single class ``giants'', because we are only interested in separating the late-type dwarfs from these other luminosity classes. We consider a star as a dwarf if it has $\wp(\text{dwarf}\vert \log g)>0.5$, and likewise a giant if $\wp(\text{dwarf}\vert \log g)\leq0.5$; the thresholds are tabulated in Table~\ref{tab:t1}. In this way, the late type subsample ($0.5<(J-K)\leq0.9$) comprises 1752 ``dwarfs'' (20.3\%) and 6882 ``giants''.

\section{The SVM Classification\label{sec:s3}}
In the previous section, we have classified the late-type stars having spectroscopic $\log g$ estimates with the help of an unsupervised mixture model classification. In this section, our goal is to use these stars to investigate whether it is possible to classify the ones that lack $\log g$ estimates into dwarfs and giants based on some other features by a supervised machine-learning algorithm. Before we decide on which features to use, we face the problem of selecting an appropriate training set of the two classes. One idea would be to cut the $\log g$-distribution at, e.g, $\log g=3.5$ and 4.0, respectively, to distinguish dwarfs from giants. Although such a choice would account for the \emph{mean} $\log g$ uncertainty of 0.5 dex \citep{zwi08} in some sense and create two class samples that could be considered as pure, it assumes totally correct $\log g$ values and can not account for the $\log g$ uncertainty of the \emph{individual} stars. In addition, the exact choice of the cutoffs is somewhat arbitrary and not directly obvious from the $\log g$-distributions (Figure~\ref{fig:f1}). Finally, by leaving out a ``gap region'' in the $\log g$-distribution one looses information from the stars that might be scattered into this region due to their $\log g$ uncertainties. 

A solution to this problem is to work in a fully probabilistic framework. Although for the classification of stars with unknown $\log g$ estimates we still need to assign a ``true'' class to each star we put into our training set, we could still weight them during the classification according to their $\log g$ values and hence their probability $\wp$ that they truly belong to their assigned class.

We therefore assign a weight $w$ to each star in a given class by mapping its probability for that class onto the interval $[0,1]$:\footnote{Although in principal weights could range from 0 to $\infty$, we choose this scaling because it puts sufficient weight onto stars with higher class probabilities and makes conversions between $\wp$ and $w$ easier to interpret.}
\begin{equation}\label{eq:e2}
	w=\begin{cases}
	   2\;\bigl(\wp(\text{dwarf}\vert\log g)-0.5\bigr) & \text{if} \wp(\text{dwarf}\vert\log g)>0.5\\
	   2\;\bigl(1-\wp(\text{dwarf}\vert\log g)-0.5\bigr) & \text{if} \wp(\text{dwarf}\vert\log g)\leq0.5.
	   \end{cases}
\end{equation}

For the machine learning algorithm we choose a SVM, because it is highly adaptable and capable of dealing with a multidimensional feature space. The SVM identifies a decision boundary in a multidimensional space based on a training set containing examples of two or more classes and assigns class probabilities to each object according to its distance from the decision boundary. We use a Java-coded implementation of LIBSVM \citep{cha01}, which is available online\footnote{\texttt{http://www.csie.ntu.edu.tw/$\sim$cjlin/libsvm/}}. We adopted the tool ``Weights for data instances'' written by Ming-Wei Chang and Hsuan-Tien Lin into our Java code.\footnote{\texttt{http://www.csie.ntu.edu.tw/$\sim$cjlin/libsvmtools/\#15}} This tool allows the user to weight the input data by creating an individual penalty parameter $C$ for each data point. In our case, we use the weights $w$ obtained from equation~\eqref{eq:e2} to weight each dwarf and giant in the training set. The training set itself is constructed from the subset of late-type stars with available $\log g$ estimates -- the 1752 ``dwarfs'' and the 6882 ``giants'' with $0.5<(J-K)\leq0.9$. 

\subsection{Choosing Appropriate Features for the Dwarf-Giant Classification}\label{sec:features}
\begin{figure}[ht!]
\epsscale{0.9}
\plotone{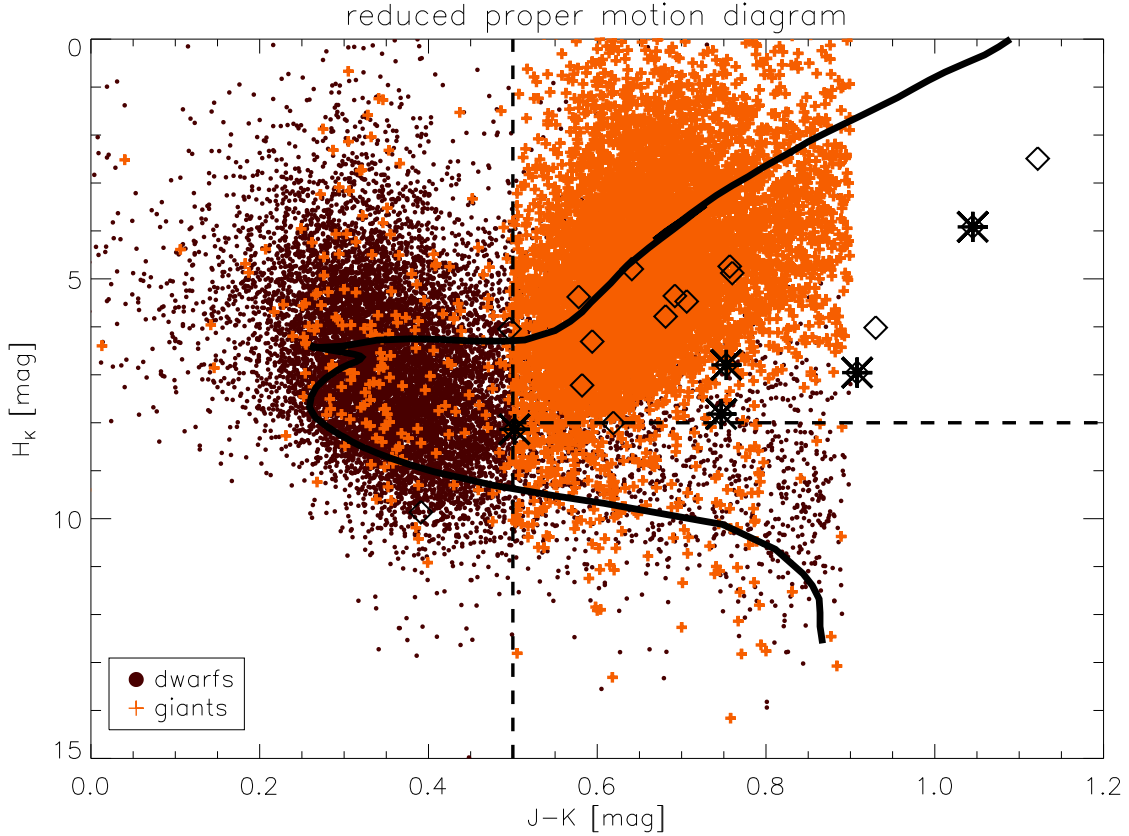}
\caption{Reduced proper motion diagram for stars in the color range $0<(j-K)\leq0.9$ that have been classified based on their spectroscopically derived $\log g$ values and the mixture models shown in Figure~\ref{fig:f1}. The black solid line is a fiducial sequence from \citet{bon04} chosen to represent a typical thin disc population. The black dotted lines are the approximate separations between K-M dwarfs and giants made by \citet{sea08}. The black star symbols denote members of the putative KFR08 stream; the black diamonds are members in the most significant ($\sigma>2$) region of the stream in Fig.~10 of \citetalias{kle08}. Note that four of the putative stream members lie in the color range $(J-K)>0.9$, where we do not try to classify stars.\label{fig:f3}}
\end{figure}

We now explore which features are generally suited to distinguish dwarfs from giants. One feature that has often been used to separate dwarfs from giants is the reduced proper motion. Figure~\ref{fig:f3} shows the ($J-K$) reduced proper motion diagram for all dwarfs (black dots) and giants (red dots) in the combined early- and late-type subsamples \citep[compare to Fig.~13 of][]{sea08}. We do not show stars with no given or improbably high (>1000 mas/yr) proper motions. This leaves 1743 dwarfs and 6877 giants in the late-type subsample and 10756 dwarfs and 208 giants in the early-type subsample, respectively. In the early-type subsample, we consider a star as a giant if it lies outside the $2\sigma$ region of the Gaussian curve in the upper left panel of Figure~\ref{fig:f1}. Note that we make no distinction between dwarfs, turnoff-stars and subgiants in the early-type subsample and include these stars in Figure~\ref{fig:f3} only for illustration, because we not use them in the SVM classification. The reduced proper motion, $H_{K}$, is
\begin{equation}\label{eq:e3}
	H_{K}=5+5\log\mu+K=5\log v_\bot+M_{K}-3.379,
\end{equation}
where $\mu$ is the total proper motion (in arcsec/year), $v_\bot$ the total transverse space velocity (in km s$^{-1}$) and $M_{K}$ the absolute magnitude in the $K$-band. Like in Figure~13 of \citet{sea08}, we show the solar-scaled Padova 2MASS isochrone \citep{bon04} as a representation of a thin disk population (Z=0.019, 2.5 Gyr, $v_\bot=60$ km s$^{-1}$\footnote{The $v_\bot=30$ km s$^{-1}$ in \citet{sea08} is a typographical error (Seabroke, private communication).}). 

Assuming that the transverse velocities of dwarfs and giants of the same spectral type are equally distributed, the reduced proper motion is a proxy for the absolute magnitude. In Figure~\ref{fig:f3}, we further show the separation lines between K-M dwarfs and giants adopted by \citet{sea08} based on the dwarf-giant bifurcation in their Figure~13. We also overplotted the 19 members of the stellar stream that we have found in our previous analysis of DR1 data (the `KFR08' stream): the blue diamonds denote members that lie in the region of phase space where the significance in Figure~10 of \citetalias{kle08} is greater than 2 (at $V\approx-160$ km s$^{-1}$), while the green stars are the other putative members that lie in a still overdense region localized by $-180$ km s$^{-1} < V < -140$ km s$^{-1}$ \citepalias[see][Section~4.3.]{kle08}. The kinematic separation implies that we classify many of the dwarfs as giants, most notably in the color range $0.8<(J-K)\leq0.9$ (see also Figure~\ref{fig:f1}). Also, we now realize that a considerable proportion of the putative stream stars are probably giants, and a more detailed analysis reveals that they have been projected into the same part of phase space because of their similar large radial velocities.

It is obvious from Figure~\ref{fig:f3} and Figure~13 in \citet{sea08} that the reduced proper motion, $H_{K}$, might be a way to separate dwarfs and giants in the late-type subsample. However, classifying stars in this way introduces a kinematic bias, which is undesirable in most applications. A dwarf-giant classification method that is particularly interesting for our study has been described by \citet{bil06}. These authors used linear decision boundaries in the space spanned by the apparent $V$ and 2MASS magnitudes, $(V,J,H,K)$, to separate FGK dwarfs (selected by $\log g>4$) from K giants ($log g<3$). As pointed out by \citet{maj03}, dwarfs and giants are degenerate in the infrared colors alone, except for the latest types ($J-K\gtrsim0.85$) where opacity effects play a role. We therefore expect to need at least one optical band in order to obtain good classification results.

\begin{figure}[ht!]
\epsscale{1.0}
\plotone{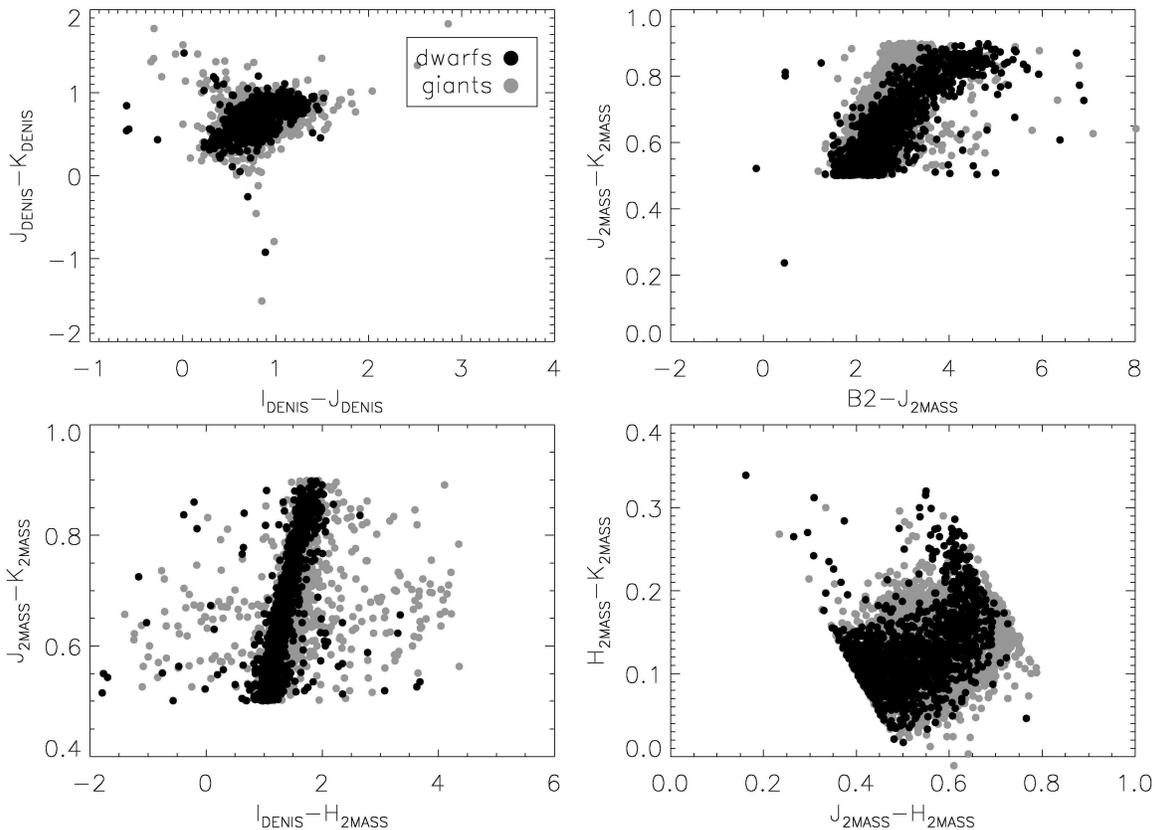}
\caption{Color-color diagrams of 1304 dwarfs (black dots) and 5488 giants (gray dots) in our training set. All stars shown occupy the color range $0.5<(J-K)\leq0.9$ and therefore are mostly spectral types K and M. \label{fig:f4}}
\end{figure}

Motivated by the work of \citet{bil06}, we show in Figure~\ref{fig:f4} the position of the two classes of our late-type subsample in different color-color diagrams. Because only a small fraction of stars have available photometry in the Tycho $V_T$ band, which is the closest available band to $V$, we choose to use the photographic USNO-B $B2$ and $R2$ bands instead.\footnote{The $V$ band ($\lambda_\text{eff}=551$ nm, FWHM=88 nm) lies between $B2$ and $R2$, which span the wavelength range of 385-540 nm and 590-690 nm, respectively \citep{mon03}.} In addition, we also make use of the DENIS $IKJ$ bands. These constraints on available photometry reduce our sample to 1305 dwarfs and 5488 giants. The diagrams already suggest that a classification of dwarfs and giants based on these colors might be difficult, because the dwarfs are distributed on top of the giants and do not occupy separate regions. This highlights the advantage of using a SVM classifier, namely that it has non-linear decision boundaries and that the weighting enables us to account for uncertainties in the training data. The next section describes the outcome of our classification attempts.

\subsection{Evaluating the Classifier Performance}
\label{sec:performance}
In this section, we examine the performance of the SVM classifier according to the use of different combinations of the following input features: $(I,J,K)_\text{DENIS}$, $(J,H,K)$\footnote{$J$ and $K$ magnitudes without index refer to 2MASS, while for DENIS magnitudes we explicitly add an index.}, $(B2,R2)$ and $H_{K}$. The choice of these features is motivated by the figures~\ref{fig:f3} and \ref{fig:f4} and the work of \citet{bil06}. It leaves us 1305 dwarfs and 5488 giants which we can use for training and testing. Through various tests, we found out that feeding the classifier with apparent magnitudes instead of colors yields statistically similar results, which shows that it is able to ``learn'' the right combinations of magnitudes by itself.

Generally, it is not a priori clear how the imbalance in the class distribution will affect a classifier. It has been shown by \citet{wei03} that the natural distribution not always performs best, while a balanced distribution in general performs well.\footnote{``Performing well'' in this context means achieving a high completeness and low contamination; see below for the definition of completeness and contamination.} Although these authors used decision trees for two-class classification, which are expected to be more sensitive to class imbalances than SVMs \citep{jap02}, it has been recommended that a SVM classifier should be trained on roughly
equal numbers of objects in each class so that it can properly learn the class distributions or boundaries \citep{bai08}. We therefore decide to use a balanced training set and use the formalism of \citeauthor{bai08} to adjust the classifier output probabilities according to the probabilities we a priori expect for each class. For the latter we adopt a simple prior given by the relative fractions of dwarfs and giants that result from our mixture models; these are 20.3\% dwarfs, 79.7\% giants. In this context we emphasize that for dwarfs and giants the priors strongly depend on the apparent magnitude limits of the survey. In our case, the stars have been drawn from the {\it Tycho}-2 and SuperCOSMOS catalogs in the Cousins $I$-band magnitude range $9<I<12$ \citep{stei06}. If RAVE reached fainter magnitudes, the fraction of giants would decrease, thereby requiring less corrections for class imbalances.  

In the machine learning literature there exist various measures of a classifier's performance \citep[e.g.][Section~4.3]{wei03}. The principle is to split the training set, for which the true classes $i$ are known, into a training and a test set and in this way assess the performance of the classifier. We can do the classification in two ways: either we assign an object to the (output) class $j$ for which its probability is highest (in our case $>0.5$); or we set a probability threshold $P_t$ and assign an object to class $j$ only if its probability for belonging to that class is higher than $P_t$. The former approach is best evaluated in the form of a confusion matrix giving true positive, false negative, false positive and true negative rates for the positive class (dwarfs) \citep[see Table~\ref{tab:t2} and][Section~4.3]{wei03}. The latter approach, however, is more suited for our purpose of building a pure dwarf sample, because it allows a more confident classification. Note that the higher the threshold, the lower the sample contamination, but the smaller its completeness. The completeness measures how many percent of objects in the test set with true class $i$ are contained within the output sample for class $j=i$. The contamination for class $j$ gives the percentage of objects in the output sample for class $j$ that are truly of other classes \citep[see also equation~(1) of][]{bai08}.

\begin{figure}[ht!]
\epsscale{1.0}
\plotone{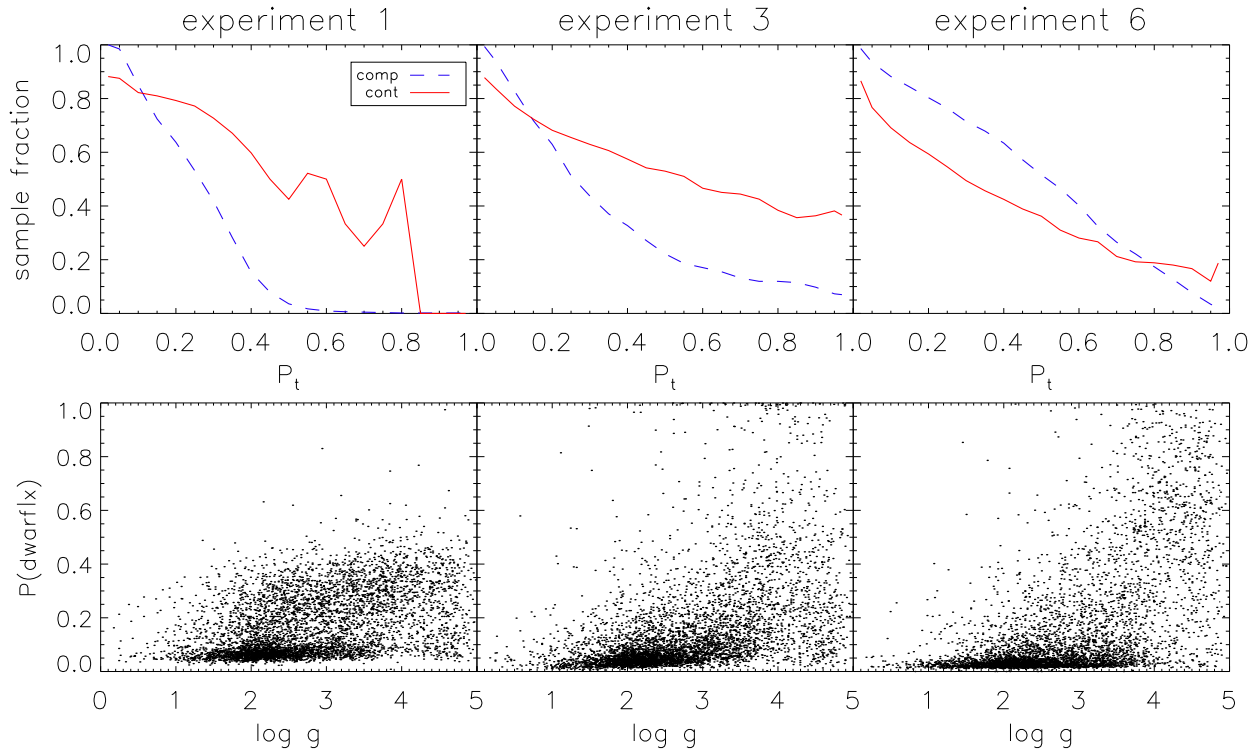}
\caption{\textit{Upper panels}: Completeness and contamination of the dwarf output sample against $P_t$ for experiments 1, 3 and 6. \textit{Lower panels}: Posterior probability of objects in the labeled test set for belonging to class ``dwarf'' given the feature vector {\bf x} vs. the $\log g$ estimate for the same experiments. Components of {\bf x} can be looked up in Table~\ref{tab:t3}. \label{fig:f5}}
\end{figure}

We have employed both approaches for different feature vectors {\bf x} that contain different combinations of the input features. We randomly select roughly half of the dwarfs and approximately the same number of giants into a new training set and use the rest of the sample to build a (labeled) test set. Although our training set weights each star according to its probability for truly belonging to the assigned class, the performance measures do not take into account these weights, only class labels. If we put stars into the test set independent of their weights, the classifier performance measures are affected by the stars which have small weights (meaning their class assignment is uncertain), but are taken to have true known classes. To investigate this effect, we build test sets in two ways: {\it in the first case}, we only put stars with weights $w>0.8$ into the test set, corresponding to class probabilities $\wp>0.9$.\footnote{This is equivalent to cutting the $\log g$-distribution of the late-type stars at $\log g=3.49$ and $\log g=2.70$, respectively, to distinguish dwarfs from giants.} {\it In the second case}, we put no constraints on the weights. 
We resample the training and testing sets in this way 100 times, each time computing the confusion matrix and completeness and contamination in the output dwarf sample for a range of different threshold probabilities $P_t$. Finally, we take the mean values and standard deviations of these performance measures to compare the classifier performance between different input features. Our experiments are summarized in Table~\ref{tab:t3} in order of decreasing contamination at a threshold $P_t=0.7$, since contamination is the most important measure for the purity of the dwarf output sample.\footnote{The choice of $P_t=0.7$ will be justified later in Figure~\ref{fig:f5}, because at this threshold the drop in contamination starts to be more shallow.} Note that the case in which the test sets include stars with small weights can be treated as the worst case which gives a lower limit on accuracy and an upper limit on the contamination. 

The outcome of the experiments have similar trends for both cases of test sets. The infrared bands alone seem to hold no valuable information about how to draw a decision boundary. This underlines the argument that a dwarf-giant separation from near-infrared bands alone is only possible for $J-K\gtrsim0.85$ \citep{maj03}. Because the SVM is trained by maximizing accuracy, it basically classifies every star as a giant. This also explains the very large standard deviation of the contamination. Adding the USNO-B $B2$ and $R2$ bands (experiments 2 and 3) improves the performance measures in a way that implies that the SVM has learned a decision boundary that separates at least a small fraction of dwarfs from the giants. It is likely that a better separation is hindered by the relatively large uncertainties of USNO-B apparent magnitudes. The best performance in terms of accuracy, completeness and contamination is obtained when 2MASS magnitudes and reduced proper motions are combined, independent of adding other photometric bands. Generally, inclusion of reduced proper motions in the feature space yields significantly higher accuracy and completeness. However, besides the fact that selecting stars based on reduced proper motions introduces a kinematic bias, the contamination of the output dwarf sample is still too high for our purpose of building a pure dwarf sample.

We give a qualitative interpretation of the trends described above in Figure~\ref{fig:f5}. In the upper panels we show the trend of completeness and contamination with probability threshold for the dwarf sample for one realization of experiments 1, 3 and 6, where in each case we had identical test sets with no constraints on the weights (case 2). We see that we can hardly push the contamination below $\sim20\%$, even if we adopt thresholds $P_t>0.7$. 

The lower panels of Figure~\ref{fig:f5} show the posterior probabilities for class ``dwarf'' given the feature vector {\bf x}, $P(\text{dwarf}\vert{\bf x})$. In experiment 1 there is only a very weak trend for stars to get higher probability for class ``dwarf'' with increasing $\log g$; in addition, the probabilities stay below 0.5 for most of the stars even in the high-$\log g$ regime. Adding $B2$ and $R2$ to the infrared 2MASS magnitudes (experiment 3) improves the classification results in the sense that now the aforementioned trend is more pronounced with a non negligible fraction of stars with $\log g\gtrsim4$ having probabilities for class ``dwarf'' higher than 0.5. Finally, adding reduced proper motions to the feature vector transfers even more of the stars with $\log g\gtrsim4$ that had $P(\text{dwarf}\vert x)<0.5$ in experiments 1 and 3 to probabilities higher than 0.5.

We conclude from our experiments that we can not use the photometric data and/or reduced proper motions of our late-type subsample to classify sources in the RAVE catalog that lack $\log g$ estimates, if we wish to obtain an output sample with contamination below 5--10\%. Making use of reduced proper motions and choosing a high threshold probability, we could eventually lower the contamination to 15-20\%, but at the expense of a low and rapidly falling completeness (Figure\ref{fig:f5}). This means that, at least for RAVE data, it is a very good and essential thing that future data releases will contain spectroscopically derived stellar parameters. However, several successful, albeit non-probabilistic, dwarf-giant separations based on photometry alone have been described in the literature \citep[e.g.][]{bek52,maj03,bil06,teig08}. We believe that the classification of stars without spectra will remain very important
for ongoing and future large-scale surveys such as PanStarrs \citep[$grizy$ bands, see Fig.~1 in][]{stu07} and LSST \citep[$ugrizy$ bands, see Fig.~4 in][]{ive08}. The accurate photometry of these surveys, spanning the optical range from $\sim350$ nm ($u$-band) to $\sim1 \mu$m ($y$-band), combined with infrared photometry from, e.g., 2MASS ($\sim$1.2$\mu$m--2.2$\mu$m) should yield much better results than in our case, where we had to rely on a narrow color range and USNO-B photometry derived from photometric plates with large uncertainties. The use of machine-learning techniques such as the SVM used by us would then allow for fully probabilistic classification with the advantage that additional knowledge (priors, other classifiers) could be easily incorporated.

In the next section we repeat our kinematic stream search from \citetalias{kle08}, for which a pure sample of dwarfs is needed. Due to the unsatisfactory classification results obtained in this section, we will only make use of the stars that we classified as dwarfs based on their $\log g$ estimates and the mixture models in \S~\ref{sec:s2}.

\begin{deluxetable}{c|cc}
\tablewidth{0pt}
\tablecaption{Confusion matrix\label{tab:t2}}
\tabletypesize{\normalsize}
\tablehead{
\colhead{} & \colhead{dwarf} &  \colhead{giant}
} 
\startdata
			DWARF & TP & FN\\
			GIANT & FP & TN\\
\enddata
		
\tablecomments{The fields TP, FN, FP and TN denote the true positive, false negative, false positive and true negative rates for the dwarf sample, respectively.} 
\end{deluxetable}

\begin{deluxetable}{c|r|c|c|c|c|c|c|c|}
\tablewidth{0pt}
\tablecaption{Results of our experiments for two cases of test sets (see text)\label{tab:t3}}
\tabletypesize{\scriptsize}
\tablehead{
\colhead{No.} & \colhead{used features {\bf x}} &  \colhead{TP} & \colhead{FN} & \colhead{FP} & \colhead{TN} & \colhead{accuracy} & \colhead{comp$_{0.7}$} & \colhead{cont$_{0.7}$}\\
\colhead{ } & \colhead{ } &  \colhead{(\%)} & \colhead{(\%)} & \colhead{(\%)} & \colhead{(\%)} & \colhead{(\%)} & \colhead{(\%)} & \colhead{(\%)}
} 
\startdata
			(1) & $(J,H,K), (I,J,K)_\text{DENIS}$ & \tc{4.0}{4.0} & \mt{96.0}{96.0} & \mt{0.5}{0.7} & \mt{99.5}{99.3} & \mt{51.8\pm0.8}{51.7\pm0.7} & \mt{0.7}{0.7} & \mt{\bm{31.9\pm24.4}}{\bm{35.6\pm23.1}}\\
			(2) & $(I,J,K)_\text{DENIS}, B2, R2$ & \tc{15.6}{13.7} & \mt{84.4}{86.3}  & \mt{2.3}{2.3}  & \mt{97.7}{97.7}  & \mt{56.7\pm1.3}{55.7\pm1.3}  & \mt{8.5}{7.1}  & \mt{\bm{58.0\pm8.1}}{\bm{52.0\pm6.0}}\\
			(3) & $(J,H,K), B2, R2$ & \tc{22.5}{20.3}  & \mt{77.5}{79.7}  & \mt{2.3}{2.8}  & \mt{97.7}{97.2} & \mt{60.1\pm1.6}{58.8\pm1.4}  & \mt{14.1}{11.5}  & \mt{\bm{40.1\pm5.0}}{\bm{41.6\pm5.6}}\\
			(4) & $(J,H,K), B2, R2, H_K$ & \tc{56.9}{54.2} & \mt{45.8}{43.1}  & \mt{2.9}{4.3}  & \mt{95.7}{97.2}  & \mt{77.1\pm1.4}{74.9\pm1.1}  & \mt{40.7}{37.6}  & \mt{\bm{27.4\pm3.2}}{\bm{32.0\pm2.3}}\\
			(5) & $(J,H,K), (I,J,K)_\text{DENIS}, B2, R2, H_K$ & \tc{59.6}{56.4}  & \mt{40.4}{43.6}  & \mt{3.2}{4.8}  & \mt{96.8}{95.2}  & \mt{78.2\pm1.4}{75.8\pm1.2}  & \mt{43.4}{39.4}  & \mt{\bm{27.8\pm3.4}}{\bm{32.0\pm2.5}}\\
			(6) & $(J,H,K), H_{K}$ & \tc{58.5}{55.6} &   \mt{41.5}{44.4} &  \mt{3.1}{4.7} & \mt{96.9}{95.3} & \mt{77.7\pm1.2}{75.4\pm1.0} & \mt{42.7}{37.3} & \mt{\bm{27.1\pm2.9}}{\bm{31.2\pm2.5}}\\
\enddata
		
\tablecomments{For each experiment, the outcomes for case 1 are given in the top line, for case 2 in the bottom line. {\bf x} denotes the feature vector. TP, FN, FP and TN are defined in Table~\ref{tab:t2}. Classification accuracy is defined as the fraction of correctly classified objects, i.e., $\frac{\text{TP+TN}}{\text{TP+TN+FP+FN}}$. The columns comp$_{0.7}$ and cont$_{0.7}$ denote the completeness and contamination of the dwarf output sample when using a threshold probability $P_t=0.7$. All values are given as the means obtained from 100 resamplings of training and test sets. For the accuracy and contamination we also quote the standard deviations.} 
\end{deluxetable}

\section{Application\label{sec:s4}}
In the previous section we have shown that we can not make use of the RAVE data that lack $\log g$ estimates, because their classification based on the available photometry and/or reduced proper motions is not reliable enough. We therefore decide to join together the ``dwarfs'' from the early and late type subsamples that we classified based on the mixture models in \S~\ref{sec:s22}. For these stars, we expect that we can use the distance estimates from \citetalias{bre10} with higher confidence than if we would simply use all available stars independent of their luminosity class. The reason is that the stellar isochrones from which \citetalias{bre10} derived their distances are systematically offset from the location of real giants. Although we have stated that the early type ``dwarfs'' contain an unknown fraction of subgiants and turnoff stars, we are confident that the \citetalias{bre10} distances are reliable enough for these stars. In fact, these authors suggested being careful when using their distances for stars with $\log g<3$ only, which is well below the dwarf-giant threshold in our early-type subsample (at $\log g=3.32$). There are, however, subgiants from later types with $\log g<3$ (Table~\ref{tab:t1}), which justifies their exclusion from our sample.

We used the TOPCAT Tool \footnote{http://www.star.bris.ac.uk/~mbt/topcat/} written by Mark Taylor to cross-match the \citeauthor{bre10} catalog \footnote{available from http://www.astro.rug.nl/~rave/} with our combined dwarf sample. The resulting sample includes 9696 stars. For our kinematic analysis, we further restricted the sample to stars with distances less than 500 pc, relative distance errors less than $40\%$, proper motion uncertainties less than 5 mas (in both components) and radial velocity uncertainties less than 5 km s$^{-1}$. This leaves 3508 stars among which we searched for stellar streams in the same way as described in \citetalias{kle08}. That is, we search for overdensities in the space spanned by $\bigl(V$,$(U^2+2V^2)^{1/2}\bigr)$, which are measures for the angular momentum and eccentricity of disk-like orbits passing the solar neighborhood. We applied a wavelet transform with scale and elongation parameters $a=7$ km s$^{-1}$ and $q=\sqrt{3}$, respectively, to enhance the overdensities \citepalias[for more details see][]{kle08}. To calculate their significance, we compare them against a smooth model that we derive from Monte Carlo simulations. The smooth model is a composite of three Schwarzschild distributions to represent the thin disk, thick disk and halo. There exists, however, a range of parameters of the three Schwarzschild distributions in the literature; the same is true for the thick-to-thin disk and halo-to-thin disk normalizations. Therefore, we test the significance of the overdensities against various smooth models and tabulate the results in Table~\ref{tab:t4}. In each case, we added additional velocity errors that we drew from a Gaussian distribution with the standard deviation set to the standard deviation of the velocity uncertainties in our sample (7.3 and 5.7 km s$^{-1}$ for $U$ and $V$, respectively). For every model, we drew 500 Monte Carlo samples and for each individual sample computed the wavelet transform with the same parameters that we used for the data. The significance is derived based on the $Z$-statistic, $Z=(w_{ij}-\bar{w}_{ij}^\text{MC})/\sigma_{ij}$, where $w_{ij}$ denotes the values of the wavelet transform for the data, $\bar{w}_{ij}^\text{MC}$ the mean obtained from the Monte Carlo samples and $\sigma_{ij}$ the standard deviation, in each case for bin $(i,j)$.

\begin{deluxetable}{c|cccccc}
\tablewidth{0pt}
\tablecaption{Parameters and fractions of the Schwarzschild distributions\label{tab:t4}}
\tabletypesize{\small}
\tablehead{
\colhead{} & \colhead{$\sigma_U$} &  \colhead{$\sigma_V$}  &  \colhead{$V_\text{lag}$}  &  \colhead{$f$}  & \colhead{$Z_\text{KFR08}$} & \colhead{$P_\text{min}$}\\
\colhead{} & \colhead{(km s$^{-1}$)} & \colhead{(km s$^{-1}$)} & \colhead{(km s$^{-1}$)} & \colhead{(\%)} & \colhead{} & \colhead{}
} 
\startdata
			model 1 & (30,60,150) & (19,40,95)  & (-10,-40,-220)  & (14,0.5)   & 2.2   & 0.09\\
			model 2 & (30,60,150) & (19,40,95)  & (-10,-40,-220)  & (14,1.0)  & 1.6  &  0.22\\
			model 3 & (39,63,150) & (20,39,95)  & (-12,-51,-220)  & (14,0.5)  & 1.5 &  0.24\\
			model 4 & (39,63,150) & (20,39,95)  & (-12,-51,-220)  & (14,1.0)  & 1.2  &  0.33\\
			model 5 & (31,59,141) & (18,48,92) &  (-10,-38,-200) & (12,0.5) & 1.2 &  0.33\\
			model 6 & (39,63,150) & (20,39,95)  & (-12,-40,-220)  & (14,1.0)  & 1.7  &  0.20\\ 
\enddata
		\scriptsize
\tablecomments{All velocities are given in the order thin disk, thick disk and halo, respectively. The normalizations $f$ are given in the order (thick disk,\,halo). $Z_\text{KFR08}$ denotes the $Z$-score for the KFR08 stream, calculated as the mean of the two bins centered at $(V_\text{az},V_{\Delta\text{E}})=(-143,217)$ km s$^{-1}$ and $(-141,215)$ km s$^{-1}$. $P_\text{min}$ is the minimum posterior probability for the hypotheses that the KFR08 overdensity is compatible with the smooth model, obtained by calibrating the $Z$-score (i.e. $p$-value) such that an interpretation as minimum Bayes factor or minimum posterior probability is justified and assuming a prior of 0.5 for the smooth model \citep[see e.g.][]{held10}.} 
\tablenotetext{1}{Chosen to give a reasonable fit to the overall velocity distribution of our sample. The normalizations agree with those given by \citet{jur08} and \citet{sou03}.}
\tablenotetext{2}{As model 2, but fraction of halo stars increased to 1\%.}
\tablenotetext{3,4}{Parameters for thin and thick disk taken from \citet{sou03}.}
\tablenotetext{5}{Parameters chosen as mean values from the literature: Thin disk velocity ellipsoids from \citet{sou03, all04}, thick disk velocity ellipsoids and normalizations from \citet{chi00,sie02,sou03,ari05,jur08,car10}, halo velocity ellipsoids and normalizations from \citet{chi00,jur08,smi09,bon10}.}
\tablenotetext{6}{Chosen to investigate the effect of varying $V_\text{lag}$ for the thick disk. Compare to model 4.}
\end{deluxetable}

\begin{figure}[ht]
\epsscale{1.0}
\plotone{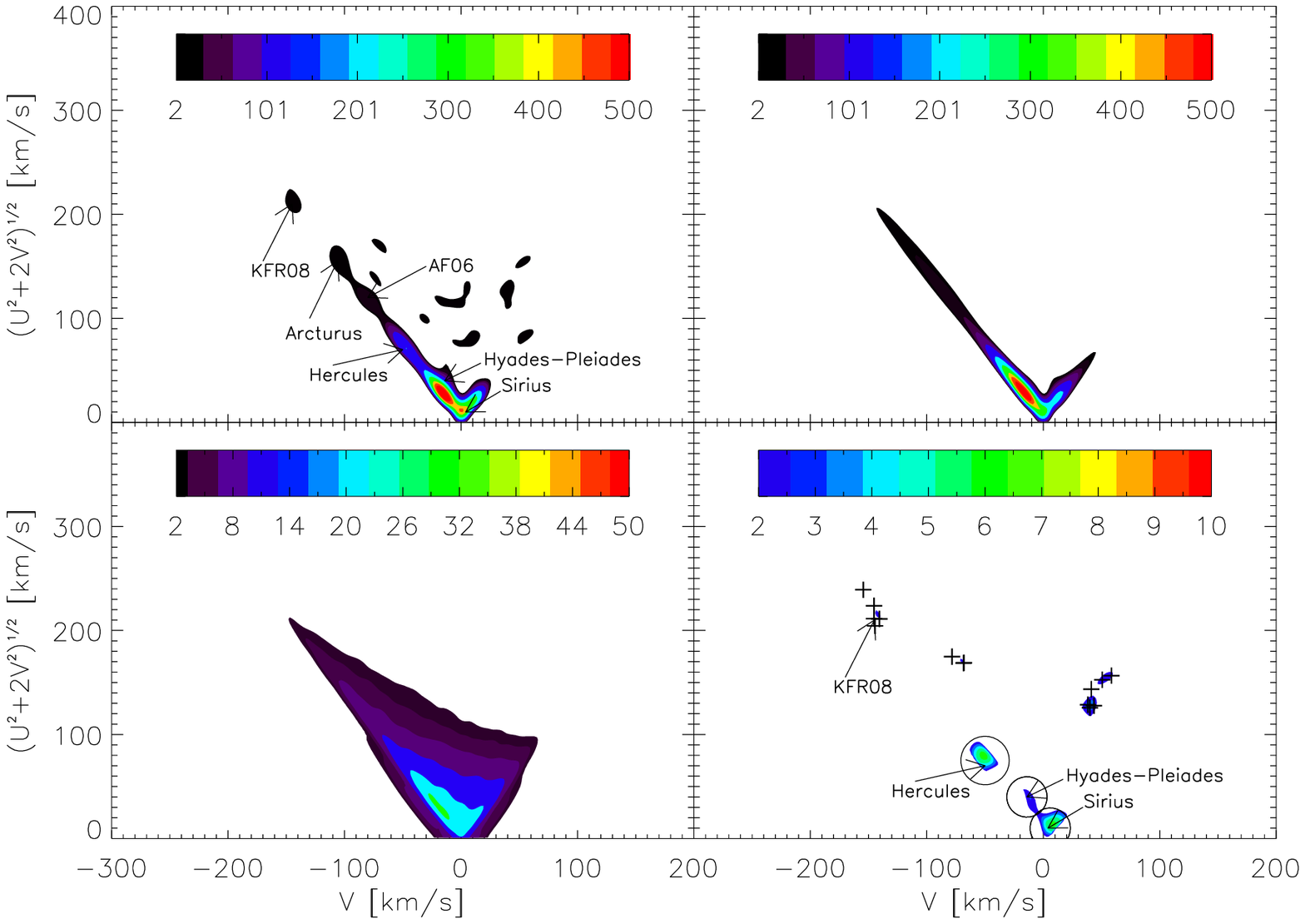}
\caption{\textit{Upper left panel}: Contours of the wavelet transform of our sample's distribution in $\bigl(V$,$((U^2+2V^2)^{1/2})\bigr)$. The same bin width and skewed Mexican hat kernel function as in \citetalias{kle08} have been used. \textit{Upper right panel}: Mean value of the wavelet transform of 500 Monte Carlo samples drawn from a Schwarzschild distribution with the ``model 1'' parameters given in Table~\ref{tab:t4}. \textit{Lower left panel}: Standard deviation of the wavelet transform of the 500 Monte Carlo samples. \textit{Lower right panel}: Significance map for our sample, showing only contours where the overdensities are $\geq2$ standard deviations above the smooth model mean. The thick crosses mark the position of stars from the KFR08 stream and those associated with the two features at $V>30$ km s$^{-1}$ (see text). The color bars give the values of the wavelet transform or the significance levels, respectively, on a linear scale.\label{fig:f6}}
\end{figure}

We demonstrate this principle of deriving significance levels for all overdensities on the basis of model 1 and Figure~\ref{fig:f6}. The upper left panel displays the distribution of our sample stars in $\bigl(V$,$(U^2+2V^2)^{1/2}\bigr)$ after applying the wavelet transform in order to enhance overdense regions. We can see hints of several well-known overdensities: Hercules at $\bigl(V$,$(U^2+2V^2)^{1/2}\bigr)\approx(-50,85)$ km s$^{-1}$, the Hyades-Pleiades streams at $(-20,40)$ km s$^{-1}$ and Sirius at (+4,9) km s$^{-1}$ \citep{fam05} or the `AF06' and Arcturus streams, which occupy the region $V\approx-80$ km s$^{-1}\approx-110$ km s$^{-1}$, respectively \citep{egg96,ari06}. In addition, there is an overdensity of 5 stars in the region of the KFR08 stream at -160 km s$^{-1}<V<-140$ km s$^{-1}$. The upper right and lower left panels show the mean value and standard deviation of the wavelet transform of the 500 Monte Carlo samples, respectively. In contrast to a single Monte Carlo sample, the mean value is a very smooth distribution, against which we compare our sample. By dividing the difference between the upper left and upper right panels through the lower left panel, we obtain the significance map showing $Z$, the standardized departure from the smooth model (lower right panel).\footnote{Like in \citetalias{kle08}, we set the standard deviation equal to one in bins where it is less than one. Also, we set the value of the wavelet transform for our sample and the mean of the Monte Carlo samples to zero whenever it is less than zero, because we are only interested in overdense regions of $\bigl(V$,$(U^2+2V^2)^{1/2}\bigr)$-space.} Note that only features with $Z\geq2$ are plotted. $Z$ measures the likelihood of our data assuming a smooth Milky Way model without kinematic substructure.

First we note that we do not want to make any inferences about features at regions in the $\bigl(V$,$(U^2+2V^2)^{1/2}\bigr)$-plane, where our smooth model predicts an underdensity of stars. These are the two features with $V$ velocities between 30 and 70 km s$^{-1}$ and the tiny spot at $\bigl(V$,$(U^2+2V^2)^{1/2}\bigr)=(-70,172)$ km s$^{-1}$. Due to the fact that underdensities artificially increase the signal, we have only indicated the position of the stars associated with these overdensities with thick crosses in Figure~\ref{fig:f6}, but do not make any further inferences about their possible association with stellar streams.

From the overdensities that we consider real, the most outstanding ones in the significance map are the three well-known streams Hercules ($Z=6.5$), Hyades-Pleiades ($Z=3.1$) and Sirius ($Z=6.3$). These are highly significant for all parameter combinations covered by the models 1--6 and are believed to be dynamical streams, resulting from the perturbations that spiral arms and/or the bar induce in the local potential \citep{fam05,ant09}. At lower rotational velocities, neither the AF06 ($Z=0.8$) nor the Arcturus ($Z=0.05$) streams are significant enough to appear in the lower right panel of Figure~\ref{fig:f6}. In fact, the overdensities attributed to these streams in the upper left panel are fully consistent with arising by chance in any of our smooth Milky Way models given in table~\ref{tab:t4}.

Finally, we detect a small signal of the KFR08 feature in which we are mostly interested in this study. It is centered at $\bigl(V$,$(U^2+2V^2)^{1/2}\bigr)=(-142,116)$ km s$^{-1}$ in Figure~\ref{fig:f6}, which departs with $Z=2.2$ from the smooth model. In this region, the mean value and standard deviation of the wavelet transform of the 500 Monte Carlo samples ranges from 0.2--0.7 and from 2.0--2.3, respectively, implying no false amplification of the KFR08 signal. However, we found that the significance of the KFR08 feature given any other smooth Milky Way model besides model 1 is not high enough to give convincing evidence for its realness (see Table~\ref{tab:t4}). Besides the choice of the halo normalization, the rotational lag of the thick disk population seems to have a major impact on the statistical significance of the KFR08 feature (compare models 4 and 6). From all six models, we infer a mean
significance level of $Z=1.6\pm0.4$, corresponding to a (one-sided) $p$-value of $0.05^{+0.07}_{-0.03}$. The $p$-value is the probability for obtaining a $Z$-value equal to or more extreme than the observed one assuming that the KFR08 overdensity is consistent with a smooth model (the null hypothesis $H_0$). However, we wish to infer $P(H_0\vert D)$, the probability of this hypothesis given our data. We obtain this posterior probability using Bayes theorem; this requires fixing a prior probability $P(H_0)$ for $H_0$ and computing the likelihood for a second hypothesis ($H_1$), e.g. a particular model that explains the observed overdensity:
\begin{equation}\begin{split}\label{eq:e4}
P(H_0\vert D)&=\frac{P(D\vert H_0)\cdot P(H_0)}{P(D\vert H_0)\cdot P(H_0)+P(D\vert H_1)\cdot (1-P(H_0))}\\
   &=\Bigl[1+\bigl(\frac{P(D\vert H_0)}{P(D\vert H_1)}\cdot\frac{P(H_0)}{1-P(H_0)}\bigr)^{-1}\Bigr]^{-1}
\end{split}
\end{equation}
Here, $P(D\vert H_0)$ and $P(D\vert H_1)$ are the likelihoods for the null and alternative hypothesis, respectively, and the prior of the alternative has been expressed through the prior of $H_0$ (as we assume that $H_0$ and $H_1$ are mutually exclusive and complete). The ratio $\frac{P(D\vert H_0)}{P(D\vert H_1)}$ is also known as the Bayes factor; smaller values of the Bayes factor provide stronger evidence against $H_0$.
Given the marginal significance of our detection, we do not consider it worth comparing different alternative models, but only assume that the alternative hypothesis is the one which maximizes the likelihood of our data and therefore minimizes the Bayes factor. In this way, we at least estimate the minimum posterior probability for the null hypothesis that the KFR08 overdensity is consistent with a smooth model. Assuming that the alternative hypothesis is the one which maximizes the likelihood of the data implies that its test statistic, $Z_1$, is distributed such that its mean is at the observed $Z$-value.
If we further assume a normal distribution for $Z_1$, the minimum Bayes factor is given by $\exp(-0.5Z^2)$ \citep{edw63,held10}. We give the null hypothesis a 50\% prior probability (i.e. we {\it a priori} give equal chance to this or any other hypothesis). From eqation~\eqref{eq:e4}, the minimum posterior probability for the smooth-model hypotheses then comes out to be in the range 0.09--0.33. These minimum probabilities for each model are tabulated in Table~\ref{tab:t4}. Other conversions between the test statistic (or $p$-value) and minimum Bayes factors are possible and usually result in somewhat larger values for the minimum posterior of the null hypothesis \citep{held10}. We therefore conclude that although the simple smooth halo models tested here seem unlikely based on their $Z$-values, their actual probabilities do not allow us to rule them out as the cause for the observed KFR08 overdensity.

There are 5 stars that we associate with the KFR08 signal. We have marked their position in the significance map of Figure~\ref{fig:f6} with black crosses and give their velocities, distances and astrophysical parameters in table~\ref{tab:t5}. A Kolmogorov-Smirnov test of their calibrated metal abundances\footnote{The typical uncertainties for the KFR08 stars are 0.2 dex \citep{zwi08}, but do not get considered in the K-S test.} against those of all other stars in the range -220 km s$^{-1}<V<-100$ km s$^{-1}$ ($N=20$) gives a probability of 0.81 that both distributions differ. However, this too is not a significant result, and a larger number of stars would certainly help drawing definite conclusions about an eventual chemical homogeneity of putative KFR08 stream members. We note that evidence for the existence of the KFR08 stream has already been found in two independent datasets: \citet{kle09}, analyzing data from the SDSS DR7, described two features at the approximate position of the `RAVE' stream, of which one (`R1') had systematically lower [Fe/H] values than the other (`R2'). The [Fe/H] values of the `R1' stream are thereby consistent with photometric [Fe/H] estimates for KFR08 candidate stars derived by \citet{bob10}. Based on revised versions of the \textit{Hipparcos} and Geneva-Copenhagen survey catalogs, \citeauthor{bob10} identified 19 putative stream members of the KFR08 stream, comprised of dwarfs, turnoff stars and giants with relative uncertainties on the parallax of less than 15\% and characterized through $(\langle V\rangle,\sigma_V)=(-151,11)$ km s$^{-1}$ and $(\langle\text{[Fe/H]}\rangle,\sigma_\text{[Fe/H]})=(-0.7,0.3)$. A comparison with the $V$ velocities of the stars we associate with the KFR08 stream ($(\langle V\rangle,\sigma_V)=(-146,5)$ km s$^{-1}$, see Table~\ref{tab:t4}) gives a fairly good agreement, altough we note that two stars have very large uncertainties in their $V$ velocity components. By fitting the Y$^2$ isochrones to the color-magnitude diagram of their stars, \citeauthor{bob10} derived an age of 13 Gyr for the KFR08 stream. They argued that ``the homogeneity of the kinematics, chemical composition, and age of the sample stars is consistent with the hypothesis that the stream is a relic remnant of the galaxy captured and disrupted by the tidal effect of our own Galaxy'', but also hypothesized about a possible connection to the Arcturus stream, which would need greater statistics to decide about. Our sample suffers from small statistics, too, but \textit{if we assume} the KFR08 stream to exist and the five stars tabulated in Table~\ref{tab:t4} to be its members, \textit{then} the high vertical velocities for four of these stars ($\vert W\vert>50$ km/s) more likely support the idea that the KFR08 stream is comprised of halo stars rather than having a dynamical origin in the (thick) disk \citep[as suggested by][]{min09}.

\begin{deluxetable}{cc|cccccccccccc}
\tablewidth{0pt}
\tablecaption{Velocities, distances and astrophysical parameters for the five stars associated with the KFR08 feature in this study.\label{tab:t5}}
\tabletypesize{\tiny}
\tablehead{
\colhead{Object ID} &  \colhead{Target ID} & \colhead{$\log g$} & \colhead{$T_\text{eff}$} & \colhead{[M/H]}
   & \colhead{$J-K$} & \colhead{$U$} & \colhead{$\sigma_U$} & \colhead{$V$} & \colhead{$\sigma_V$} &\colhead{$W$} & \colhead{$\sigma_W$} &\colhead{$d$} &\colhead{$\sigma_d$}\\
   \colhead{ }  &  \colhead{ } & \colhead{(dex)} & \colhead{K} &  \colhead{(dex)} &  \colhead{(mag)} & \colhead{(km s$^{-1}$)} & \colhead{(km s$^{-1}$)} & \colhead{(km s$^{-1}$)} &\colhead{(km s$^{-1}$)} & \colhead{(km s$^{-1}$)} & \colhead{(km s$^{-1}$)} & \colhead{(pc)} & \colhead{(pc)}     } 
\startdata
C0622232-573246 & J062223.3-573246 & 4.74 & 6597  &  -0.10 &  0.34 & 71 & 22 & -141 & 7 & -24 & 11.6 & 466 & 142\\
T8496\_00634\_1  & J032150.0-552442  & 4.31 & 5744 & -0.22 & 0.52 & -8 & 7 & -145 & 8 & -135 & 7 & 119 & 46 \\
T4900\_01632\_1 & J093415.8-054505 & 4.46 & 5919 & -0.03 & 0.48 & -50 & 6 & -145 & 7 & 90 & 5 & 211 & 67 \\
T8023\_00143\_1 & J001548.9-450433 & 4.27 & 5847 & -0.13 & 0.43 & 89 & 30 & -145 & 57 & 59 & 20 & 311 & 118 \\
C0229387-130746 & J022938.7-130746 & 4.71 & 6916 & -0.82 & 0.36 & -97 & 8 & -155 & 44 & -135 & 7& 422 & 123\\
\enddata
\tablecomments{\footnotesize All parameters are taken from the catalog of \citet{bre10}. [M/H] denotes the calibrated metallicity estimate \citep[see also][]{zwi08}. $(U,V,W)$ are the Cartesian velocity coordinates with respect to the Local Standard of Rest and $(\sigma_U,\sigma_V,\sigma_W)$ their corresponding uncertainties. $d$ and $\sigma_d$ denote the distance from the sun and its uncertainty estimate.}
\end{deluxetable}

\section{Conclusions\label{sec:s5}}
With the release of the second RAVE data release, which contains $\log g$ values for 14,772 individual stars, we were able to create a set of stars with probabilities for belonging to the two classes ``dwarf'' and ``giant'' by fitting the $\log g$ distributions with multiple component mixture models. The estimated high fraction of giants calls for a critical re-analysis of our recent study \citepalias{kle08}. For this purpose we tested whether we could use the stars to which we had assigned class labels to classify the ones without $\log g$ estimates in order to enlarge the dwarf sample. We restricted ourselves to the color range $0.5<J-K<0.9$, because there the association between one component of the mixture model and the dwarfs was clearest. Using a SVM algorithm, we investigated which features would be suitable for that task by splitting our labeled sample into a training and test set and evaluating the classifier performance for different feature vectors. These comprised various combinations of photometric bands and reduced proper motions. As already suggested by Figure~\ref{fig:f4} and arguments made by \citet{maj03}, the classifier was not able to draw a meaningful boundary between dwarfs and giants from infrared magnitudes alone; the SVM classified nearly all the stars as giants in order to achieve the highest possible accuracy ($\sim50\%$). Adding the optical $B2$ and $R2$ magnitudes improved the classification. It is difficult to compare our result to the findings of \citet{bil06}, who separated dwarfs from giants based on their apparent $V$ and $(J,H,K)$ magnitudes alone: First, these authors do not give any performance measure of their classification. From a close inspection of their Fig.~1, it seems that the giant contamination in the dwarf sample is very small, but it should also be noted that their samples lack any intermediate stars with $3\leq log g\leq4$ that might add further confusion. Second, $B2$ and $R2$ are derived from photographic plates and posess relatively large uncertainties \citep[$\sim0.25$ mag,][]{mon03}. Third, the $V$ band is located between $B2$ and $R2$, so that our results can not be directly compared to the work of \citeauthor{bil06}. Our best results in terms of accuracy and contamination have been obtained when we used reduced proper motions in the feature vector. However, even with a classification accuracy around 75\%, the contamination in the output dwarf sample was higher than 25\% (Table~\ref{tab:t3}). We therefore conclude that photometric data for RAVE stars together with reduced proper motions alone are not adequate when the goal is to build a sample of dwarfs with minimal contamination from giants.

Being confident about the classifications we obtained from the mixture models in \S~\ref{sec:s22}, we decided to take the labeled dwarfs from both the early and late-type subsamples and combine them with the distance estimates given in the \citetalias{bre10} catalog. These stars posses high enough $\log g$ parameters to have reliable distance estimates. Finally, we only kept stars with low uncertainties on the proper motion and radial velocity ($\sigma_{\mu_\alpha}\leq5$ mas, $\sigma_{\mu_\delta}\leq5$ mas, $\sigma_{v_\text{rad}}\leq5$ km s$^{-1}$) and relative uncertainties on the distance less than 40\%.

In the resulting sample of 3508 stars, we searched for nearby stellar streams in the same way as described in \citetalias{kle08}. This requires looking for overdensities in the space spanned by $\bigl(V$,$(U^2+2V^2)^{1/2}\bigr)$, which are proxies for a nearby star's angular momentum and eccentricity. Due to the variablility of Galactic model parameters given in the literature, we repeated our analysis for five different choices of parameters (tabulated in Table~\ref{tab:t4}). We detected the well-known moving groups Sirius, Hyades-Pleiades and Hercules at high confidence levels for every choice of parameters. We also found hints for the existence of the AF06 and Arcturus streams in the data, but they turned out to be consistent with resulting by chance from a smooth distribution ($Z<1$). We found that the KFR08 stream is only marginally significant and reaches our required level of 2.0 only for one particular model (model 1). The choice of the halo normalization as well as the rotational lag of the thick disk component seem to be critical to the inferred significance level. For five stars associated with the KFR08 feature, the metallicity distribution deviates from that of the other stars in the range $-220<V<-100$ km s$^{-1}$ at 81\% confidence. We point out that the number of stars in this range, which are used in the test, is small (in total $N=25$). We also note that the $V$-velocities of two of the putative KFR08 stars posess large uncertainties (Table~\ref{tab:t5}). We therefore can neither confirm nor discard the realness of the KFR08 stream with any degree of confidence. Ulimately, one has to search for its signature in larger and/or independent data sets. Recently, evidence for its existence has been found in two independent datasets \citep{kle09,bob10}. Hopefully, stronger constraints on the realness of the `RAVE' stream can be put forward by RAVE itself as future data releases become available.    

\begin{acknowledgments}
We would like to thank the anonymous referee for valuable comments that helped to significantly improve a former version of this paper. Rainer Klement would like to thank Erik Anderson and Charles Francis for useful discussions about the RAVE stars that led to the initiation of this paper. Funding for RAVE has been provided by: the Anglo-Australian Observatory; the Astrophysical Institute Potsdam; the Australian National University; the Australian Research Council; the French National Research Agency; the German Research foundation; the Istituto Nazionale di Astrofisica at Padova; The Johns Hopkins University; the W.M. Keck foundation; the Macquarie University; the Netherlands Research School for Astronomy; the Natural Sciences and Engineering Research Council of Canada; the Slovenian Research Agency; the Swiss National Science Foundation; the Science \& Technology Facilities Council of the UK; Opticon; Strasbourg Observatory; and the Universities of Groningen, Heidelberg and Sydney. 
\end{acknowledgments}

\end{document}